\documentclass{jpsj-suppl}
\usepackage{txfonts} 

\title{
Muon spin spectroscopy in multiferroic (Cu,Zn)$_{3}$Mo$_{2}$O$_{9}$
}

\author{%
Haruhiko \textsc{Kuroe}$^{1}$, %
Kento \textsc{Aoki}$^{1}$, %
Tasuku \textsc{Sato}$^{1}$, %
Ryo \textsc{Kino}$^{1}$, %
Hideki \textsc{Kuwahara}$^{1}$, %
Tomoyuki \textsc{Sekine}$^{1}$, %
Masashi \textsc{Hase}$^{2}$, %
Ikuto \textsc{Kawasaki}$^{3}$, %
Takayuki \textsc{Kawamata}$^{3,4}$, %
Takao \textsc{Suzuki}$^{3,5}$, %
Isao \textsc{Watanabe}$^{3}$, %
Kunihiko \textsc{Oka}$^{6}$, %
Toshimitsu Ito$^{6}$, %
and Hiroshi \textsc{Eisaki}$^{6}$%
}

\inst{%
$^{1}$
Physics Division, Sophia University, %
7-1 Kioi-cho, Chiyoda-ku, Tokyo 102-8554, Japan \\
$^{2}$
National Institute for Materials Science (NIMS),%
 Tsukuba, Ibaraki 305-0047, Japan\\
$^{3}$
Advanced Meson Science Laboratory,
RIKEN Nishina Center,
Wako, Saitama 351-0198, Japan\\
$^{4}$ Department of Applied Physics, Graduate School of Engineering, Tohoku University, %
Aoba-ku, Sendai 980-8579, Japan\\
$^{5}$ College of Engineering, Shibaura Institute of Technology, %
Minuma, Saitama 337-8570, Japan\\
$^{6}$
National Institute of Advanced Industrial Science and Technology (AIST), %
Tsukuba, Ibaraki 305-8568, Japan
}

\email{kuroe@sophia.ac.jp}

\recdate{\today}

\abst{
We present the muon spin relaxation/rotation spectra 
in the multiferroic compound (Cu,Zn)$_{3}$Mo$_{2}$O$_{9}$. 
The parent material Cu$_{3}$Mo$_{2}$O$_{9}$ has 
a multiferroic phase below $T_{\rm N}$ = 8.0 K, 
where the canted antiferromagnetism and the ferroelectricity coexist.
The asymmetry time spectra taken at RIKEN-RAL pulsed muon facility 
show a drastic change at $T_{\rm N}$. 
At low temperatures 
the weakly beating oscillation caused by the static internal magnetic fields 
in the antiferromagnetic phase was observed in Cu$_{3}$Mo$_{2}$O$_{9}$ 
and the lightly (0.5\%) Zn-doped sample. 
From the fitting of the oscillating term, 
we obtain the order parameter in these samples: 
ferromagnetic moment in two sublattices of antiferromagnet.
In the heavily (5.0\%) Zn-doped sample, 
the muon-spin oscillation is rapidly damped. 
The frequency-domain spectrum of this sample suggests 
a formation of magnetic superstructure.
}

\kword{$\mu$SR, quantum spin system, multiferroic material}

\begin{document}
\maketitle

\section{Introduction}
Physics on quantum magnet 
has been attracted much attention.
One of the most interesting effects is 
a quantum phase transition (QPT).
It has been studied 
in many kinds of quantum magnets, 
such as an interacting antiferromagnetic (AFM) spin dimer system 
described well by using the bond-operator theory \cite{Matsumoto2004}.
In this system, 
a pair of $S = 1/2$ spins forms a spin-singlet dimer. 
As a result of interaction between spin dimers, 
the magnetic excitation becomes dispersive.
The ground state is nonmagnetic 
and its magnetic excitations are characterized 
by a spin gap much smaller than 
the amplitude of the AFM exchange interaction 
along the path forming the spin dimer.
In the case when magnetic field or pressure suppresses the spin gap, 
the ground state becomes a mixed singlet-triplet state; 
the system becomes magnetic \cite{Matsumoto2004}.

As well as external fields, 
impurity substitution 
has a potential 
to cause QPT from the gapfull state to a gapless AFM one.
In this proceeding paper, we focus on site substituted systems.
Tl(Cu,Mg)Cl$_{3}$ system gives 
a good example \cite{Oosawa2002,Oosawa2003,FOOTNOTE}.
QPT occurs through the following process:
(i) A nonmagnetic Mg$^{2+}$ ion is 
substituted for a $S = 1/2$ Cu$^{2+}$ ion 
so that the singlet ground state of spin dimers is broken; 
(ii) As a result, a nonmagnetic impurity substitution 
induces unpaired spins with $S=1/2$; 
(iii) Impurity-induced AFM state
is formed through the interaction between the unpaired spins.
This process is known as an `order by disorder' effect.
The muon spin spectroscopy 
has been applied to this system 
as an efficient local probe 
\cite{Suzuki2009JPSJ,Suzuki2009PRB,Suzuki2010JPCS-225,Suzuki2010JPCS,Suzuki2011PRB}. 

In this study, 
we focus on QPT in (Cu,Zn)$_{3}$Mo$_{2}$O$_{9}$
induced by the site substitution.
The parent material, 
Cu$_{3}$Mo$_{2}$O$_{9}$, has a multiferroic phase, 
where the slightly canted antiferromagnetism 
and the spin-driven ferroelectricity coexist \cite{Hamasaki2008,Kuroe2011}.
From the viewpoint of magnetism, 
this compound has a unique 
distorted tetrahedral spin chain 
made from $S=1/2$ Cu$^{2+}$ ions, 
where quantum fluctuations from the low dimensionality 
and geometrical spin frustration effects coexist.
The magnon dispersion curve 
obtained by inelastic neutron scattering shows that 
the spin system is regarded as 
the AFM spin dimers indirectly interacting with each other 
through the quasi-one dimensional spin system\cite{Kuroe2010,KuroeNeutron2011,Matsumoto2012}.
We will show magnetization in (Cu,Zn)$_{3}$Mo$_{2}$O$_{9}$ system 
in the following section.
The multiferroic properties in the Zn-5.0\% sample, 
including temperature and magnetic-field dependences of 
(differential) magnetization, dielectric constant, 
a electric-polarization-electric-field loop, 
and specific heat have been reported 
in our recent publication \cite{Kuroe2013}.
In this proceedings paper, 
we focus on the muon-spin relaxation/rotation ($\mu$SR) spectra 
in (Cu,Zn)$_{3}$Mo$_{2}$O$_{9}$
at zero magnetic field.

\section{Experiments}
\subsection{Samples}
\begin{figure}[tbh]
\begin{center}
\begin{minipage}[b]{0.35\textwidth}
\caption{
(a) Temperature variations of magnetization at 0.1 T 
in (Cu,Zn)$_{3}$Mo$_{2}$O$_{9}$ 
with Zn concentrations of 0\%, 0.5\%, and 5.0\%.
(b) Reduction of the low-temperature 
magnetization in Cu$_{3}$Mo$_{2}$O$_{9}$.
\label{MT}
}
\end{minipage}
\hspace{1em}
\includegraphics[width=0.5\textwidth]{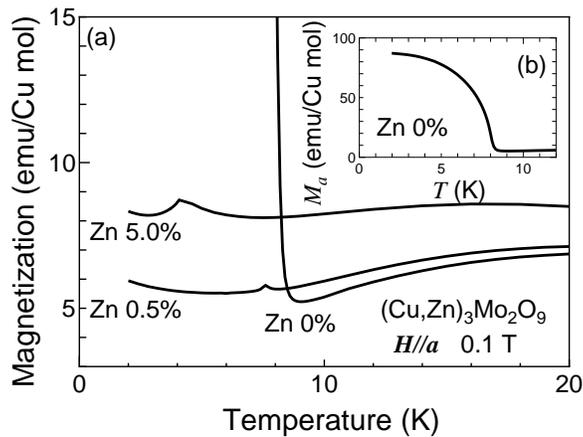}
\end{center}
\end{figure}
The single crystals of (Cu,Zn)$_{3}$Mo$_{2}$O$_{9}$ 
were prepared with the continuous solid-state crystallization method
by using an infrared furnace utilized 
for the floating-zone method \cite{Oka2011}.
This method gave single crystals 
of highly homogeneous impurity substitution 
because they were grown {\it without} melting.
We prepared platelike single crystals of Cu$_{3}$Mo$_{2}$O$_{9}$ 
and the Zn-0.5\% and the Zn-5.0\% substituted samples 
by slicing them perpendicular to the $a$ axis. 
The thickness of the sample is chosen to be 350 $\mu$m, 
which is enough to stop the muon beam.

To explain the magnetic properties in these samples, 
we show their magnetizations as functions of temperature 
in Fig. \ref{MT}. 
As shown in Fig. 1(b), 
a sharp increase of the magnetization $M_{a}$ along the $a$ axis 
with decreasing temperature below $T_{\rm N}$ was observed
in Cu$_{3}$Mo$_{2}$O$_{9}$. 
As will be discussed later in detail, 
this increment is due to the weak ferromagnetic moment 
in the canted AFM phase induced by magnetic fields.
In the Zn-0.5\% and the Zn-5.0\% samples, 
as shown in Fig. 1(a), 
small cusps were observed 
instead of the increase of magnetization due the spin canting.
Moreover, a small increase of magnetization 
probably obeying a Curie-Weiss law 
was observed at low temperatures.
The origin of it is thought to be 
the disorder effects on the distorted tetrahedral spin chains 
which include the unpaired spins on the AFM spin dimers 
induced by the nonmagnetic Zn doping.

\subsection{$\mu$SR measurements}
The $\mu$SR time spectra are measured 
using the pulsed surface muon at RIKEN-RAL muon facility, 
of which momentum and kinetic energy are 
29.79 MeV/$c$ and 4.119 MeV, respectively\cite{Matsuzaki2001}.
The single crystals were mounted on 
the high-purity silver plate 
using Apiezon N grease, 
so that the muon beam with a diameter of $\phi$25 mm 
is irradiated efficiently.
The spin polarization of implanted muons 
is parallel to the $a$ axis, 
along which the spontaneous magnetization
is observed.
This is not the direction of spin chain; 
the weak ferromagnetic components of 
the spin moments of the distorted tetrahedral spin chains 
at the center and the corner of the orthorhombic unit cell 
direct to the direction of $\pm$23.2$^\circ$ from the $a$ axis, 
respectively \cite{Hamasaki2008}.
They have both of the weak ferromagnetic components 
along the $a$ and $c$ axes, 
the latter of which are canceled in the unit cell.
The single crystals were cooled down to 0.3 K 
in a ${}^3$He cryostat with a charcoal sorption pump.
To ensure thermal contact, 
the samples were wrapped tightly 
in a silver foil (thickness 25 $\mu$m). 
The observed $\mu$SR asymmetry parameter 
$A_{\rm obs}(t)$ was defined as 
\begin{equation}
A_{\rm obs}(t) =
\displaystyle 
\frac{F(t) - \alpha B(t)}{F(t) + \alpha B(t)}
\ ,
\end{equation}
where $F(t)$ and $B(t)$ were the total muon events 
counted by the forward and the backward counters at time $t$, 
respectively.
The factor $\alpha$ calibrates 
relative counting efficiencies between them.
The $\alpha$ is obtained 
from the $A_{\rm obs}(t)$ above $T_{\rm N}$ 
under a weak transverse magnetic field, 
which is a standard method to calibrate $\alpha$.
Raw data of $A_{\rm obs}(t)$ contain 
the signals from muons 
stopped at the Ag foil and the sample holder, 
$A_{\rm Ag}$, of which value (about 6\%) 
depends on the measurement setting.
These unwanted components 
were removed by using $A_{\rm obs}(t)$ 
measured under the weak transverse magnetic field 
at temperatures much below $T_{\rm N}$, 
which is also a standard method to obtain $A_{\rm Ag}$.
We introduce a corrected asymmetry spectrum $A(t)$ 
at temperature $T$ 
as $[A_{\rm obs}(t) - A_{\rm Ag}]$ 
normalized by $[A_{\rm obs}(0) - A_{\rm Ag}]$ at 10 K, 
so that $A(t)$ at 10 K decays from $A(0)$ = 1.
Here 10 K is chosen as a temperature well above $T_{\rm N}$.

\section{Results}
\begin{figure}[tbh]
\includegraphics[width=\textwidth]{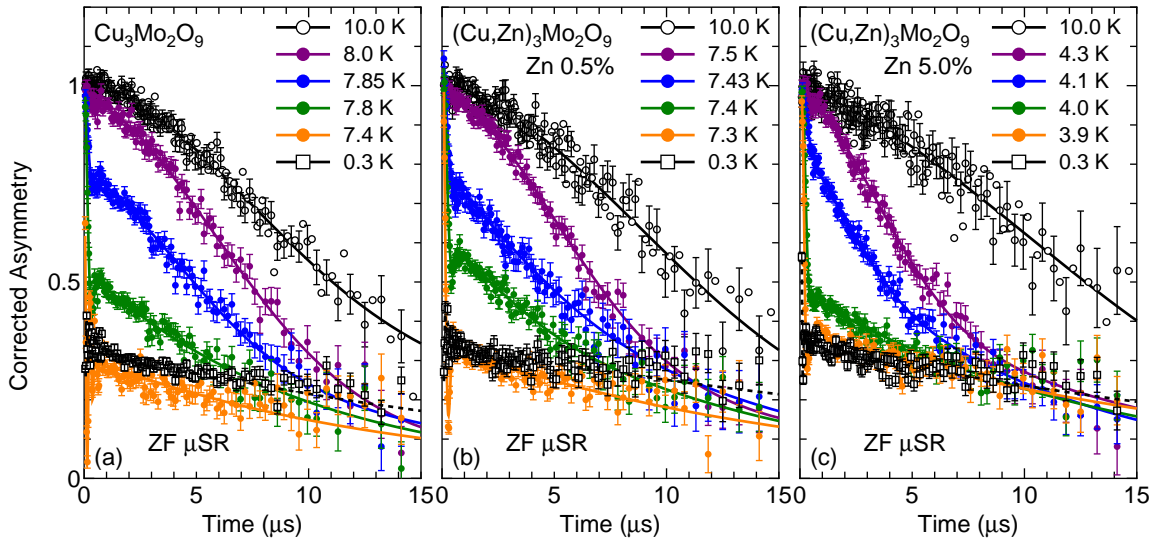}
\caption{(color online)
Muon time spectra in Cu$_{3}$Mo$_{2}$O$_{9}$ 
at various temperatures below and above $T_{\rm N}$
and those in the samples with Zn-0.5\% and Zn-5.0\% substitutions 
shown in (a), (b), and (c), respectively.
}
\label{CorrectedAsymmetry}
\end{figure}

Figures \ref{CorrectedAsymmetry}(a)-\ref{CorrectedAsymmetry}(c) 
show $A(t)$ in Cu$_{3}$Mo$_{2}$O$_{9}$ 
and the Zn-0.5\% 
and the Zn-5.0\% substituted samples, respectively.
At 10 K, well above $T_{\rm N}$, 
$A(t)$ seems to obey a Gaussian function.
We call this component the G term.
The G term appears 
as a small-$t$ part of 
the Kubo-Toyabe relaxation function 
which reproduces the muon spin relaxation 
of the paramagnetic polycrystal 
where randomly aligned static internal fields 
work at the muon sites.
Here the word `static' indicates that 
the time scale of internal field fluctuation 
is much slower than 
the time scale of $\mu$SR measurement (10$^{-6}$ -- 10$^{-5}$ sec).
Reflecting critical phenomena, 
the G term rapidly disappears 
with decreasing temperature
just below $T_{\rm N}$.

At 0.3 K, well below $T_{\rm N}$, 
$A(t)$ seems to show a exponential decay 
from $A(0)$ much smaller than the unity.
We call this term the Lorentzian term or the L term, 
in a customary way of muon spectroscopy.
The L term is observed 
through the strong collision model 
with the dynamical fluctuation of internal fields 
of which time scale is much faster than 
the time scale of $\mu$SR measurement.
The L term was observed even above $T_{\rm N}$ 
as a background of the Gaussian term.
$A(0)$ below $T_{\rm N}$ is much smaller than the unity, 
suggesting that 
there is a rapid damping term 
of which damping speed is much faster 
than the dead time of $\mu$SR measurement.
Because this component is undetectable with a pulsed muon source, 
we call this term `missing term' or the M term.

\begin{figure}[tbh]
\includegraphics[width=\textwidth]{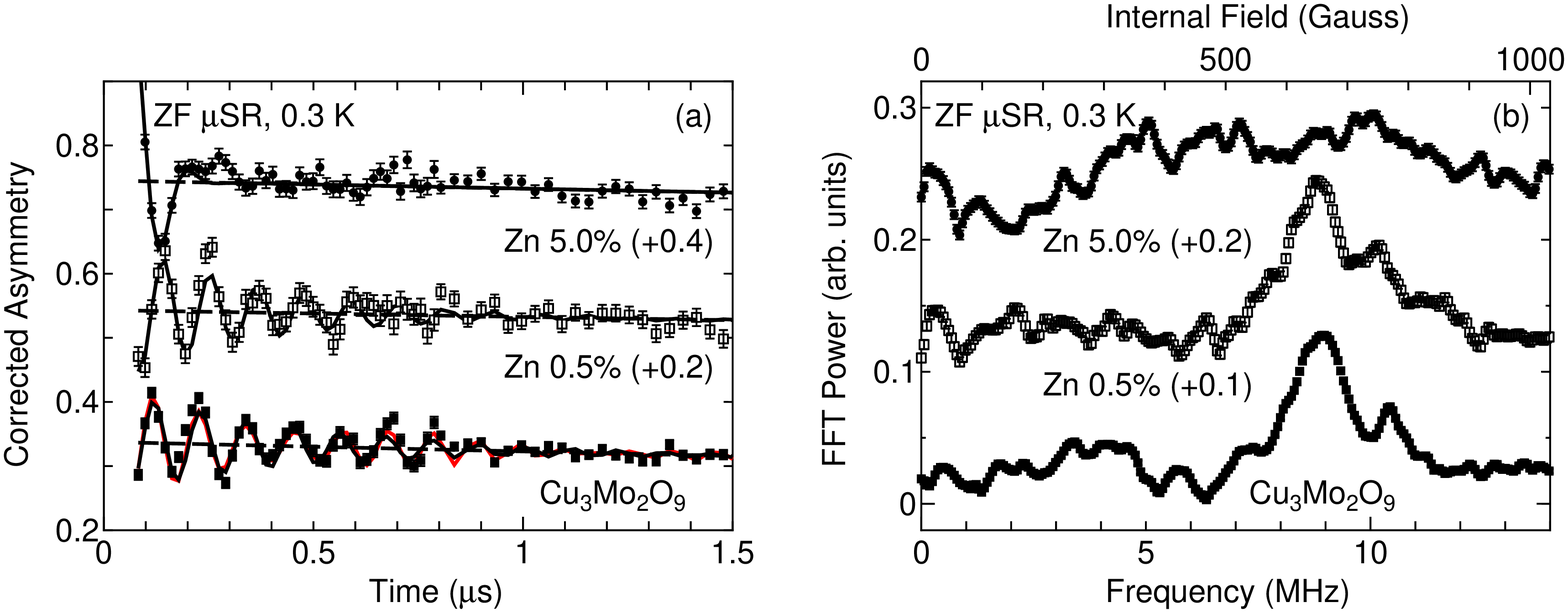}
\caption{(color online)
(a) Muon time spectra at 0.3 K in Cu$_{3}$Mo$_{2}$O$_{9}$ 
and those in the samples with Zn-0.5\% and Zn-5.0\% substitutions.
The oscillating and exponentially decaying components are 
shown by the solid and the dashed curves, respectively.
The red curve denotes the oscillating component 
obtained by using the two oscillator model, 
of which the details are given in the text.
(b) The fast Fourier transformation power spectra of oscillating terms.
}
\label{CorrectedAsymmetry_003K}
\end{figure}
As expanded in Fig. \ref{CorrectedAsymmetry_003K}(a), 
$A(t)$ contains an oscillating term.
We call it the O term.
The oscillation of $A(t)$ of Cu$_{3}$Mo$_{2}$O$_{9}$ 
in Fig. \ref{CorrectedAsymmetry_003K}(a) shows beat signals 
around 0.25 and 0.75 $\mu$s.
It was clearly shown in its fast Fourier transformation 
in Fig. \ref{CorrectedAsymmetry_003K}(b) 
as a two-peak structure with the internal fields of 667 and 760 Gauss.
The oscillating component and its fast Fourier transformation 
of the Zn-0.5\% substituted sample 
is qualitatively similar to those of Cu$_{3}$Mo$_{2}$O$_{9}$. 
The fast Fourier transformation of this sample 
seems to have a main component at 667 Gauss. 
However, the noisy spectrum prevented 
us from obtaining the detailed structure above 700 Gauss.
This is due to the noisy spectrum around $t \sim 0.7$ $\mu$s 
while the beat signals were clearly observed around 0.7 $\mu$s 
in case of Cu$_{3}$Mo$_{2}$O$_{9}$.
The possible origins of the beating oscillation in time-domain spectra 
and the multipeak structure in frequency-domain spectra 
are thought to be the phase separation and/or 
the multiple muon stopping sites.
The $A(t)$ and its fast Fourier transformation 
in the Zn-5\% sample are very different from 
those in other samples.
The damping rate of oscillation in $A(t)$ is much faster.
The internal field has a very broad spectrum
around 500 Gauss.
These indicate widely distributed internal fields.

Next, let us discuss the fitting function.
Unfortunately, there is no function 
which fits all the data in all the samples.
And then, we need to combine two components 
each of which has the Gaussian or the Lorentzian decay.
The fitting function 
of $A(t)$ above $T_{\rm N}$ in all the samples is 
\begin{equation}
A_{\rm HT}(t)
= A_{\rm G} \exp{\left[ - \left( \sigma_{\rm G} t \right)^{2} \right]}
+ A_{\rm L} \exp{\left[ - \lambda_{\rm L} t^{\mathstrut} \right]} \ ,  
\end{equation}
where the suffix G(L) denotes the quantity for the Gaussian (Lorentzian) decay
with a specific rate $\sigma_{\rm G}$ ($\lambda_{\rm L}$).
Here we introduced the condition of $A_{\rm G} + A_{\rm L} = 1$ 
for amplitude parameters $A_{\rm G}$ and $A_{\rm L}$.

Below $T_{\rm N}$, we consider 
the main component of the O term 
in cases of Cu$_{3}$Mo$_{2}$O$_{9}$ and the Zn-0.5\% sample, 
which oscillates with the frequency $\gamma H_{\rm int}$, 
where $\gamma$ and $H_{\rm int}$ are 
the gyromagnetic ratio of muon
and the internal field at the muon stopping site(s), respectively, 
which leads the strongest amplitude in fast Fourier transformations.
Reflecting the frequency domain spectrum 
in Fig. \ref{CorrectedAsymmetry_003K}(b), 
the damping factor of the oscillating term 
is chosen to be the exponentially decaying function 
and $A(t)$ is given as:
\begin{equation}
A_{\rm LT-L}(t)
= A_{\rm O} \cos{\left[\gamma H_{\rm int} t + \phi^{\mathstrut} \right]} 
            \exp{\left[ - \lambda_{\rm O} t^{\mathstrut}  \right]}
+ A_{\rm L} \exp{\left[ - \lambda_{\rm L} t^{\mathstrut} \right]} \ ,  
\label{OneOscillator}
\end{equation}
with the factor $A_{\rm M} \equiv 1-A_{\rm O} - A_{\rm L}$ 
which corresponds to the decrease of the initial asymmetry at 0.3 K;
in other words, it corresponds to the amplitude of the M term.
Here the phase factor $\phi$ is introduced phenomenologically.
One can see that this model could not reproduce 
the beat of the $\mu$SR time spectra in Fig. 3(a) 
because the dominating oscillator with a single frequency is considered.
For comparison, we introduce the two oscillators model 
where the term of 
$A_{\rm O}' \cos{\left[\gamma H_{\rm int}' t + \phi^{\mathstrut} \right]} 
            \exp{\left[ - \lambda_{\rm O}' t^{\mathstrut}  \right]}$
is added to eq. (\ref{OneOscillator}).
We show the fitting result 
by the red curve in Fig. \ref{CorrectedAsymmetry_003K}(a). 
The beating spectrum is reproduced slightly better 
when we fit it to the two oscillators model;
however, the obtained parameters of 
the dominating oscillatior are not changed 
within the experimental accuracy as listed in the following table:
\begin{table}[h]
\begin{tabular}{r|cccc}
\parbox[t]{0.2\textwidth}{
\begin{center}
Model
\end{center}
} & 
\parbox[t]{0.2\textwidth}{
\begin{center}
Corrected Initial \\ Asymmetry (\%)
\end{center}
} & 
\parbox[t]{0.12\textwidth}{
\begin{center}
$A_{\rm O}$ or $A_{\rm O}'$ (\%)
\end{center}
} & 
\parbox[t]{0.12\textwidth}{
\begin{center}
$H_{\rm int}$ or $H_{\rm int}'$ (Gauss)
\end{center}
} & 
\parbox[t]{0.12\textwidth}{
\begin{center}
$\lambda_{\rm O}$ or $\lambda_{\rm O}'$ (MHz)
\end{center}
} \\
\hline
One Oscillator Model & 
42.53$\pm$0.65& 
8.73$\pm$0.65	&
667.0$\pm$1.5	&
2.31$\pm$0.22	\\
Two Oscillators Model &
43.82$\pm$0.55&
8.54$\pm$0.53	&
665.8$\pm$1.4	&
2.14$\pm$0.19	\\
&
&
1.49$\pm$0.19	&
759.1$\pm$1.6	&
0.61$\pm$0.22	\\
\end{tabular}
\caption{
Corrected initial asymmetries, the relative amplitudes, the internal fields, 
and the damping factors obtained by using the one oscillator and 
the two oscillators models.
}
\label{Models}
\end{table}

\noindent
Hereafter, we fit the experimental data 
to the one oscillator model 
which leads the precise parameters 
within experimental error.
This model is applicable to 
the $\mu$SR time spectra 
at high temperatures and 
those in the Zn-substituted samples, 
which are slightly scattered 
due to short accumulation times.

\begin{figure}[tbh]
\includegraphics[width=\textwidth]{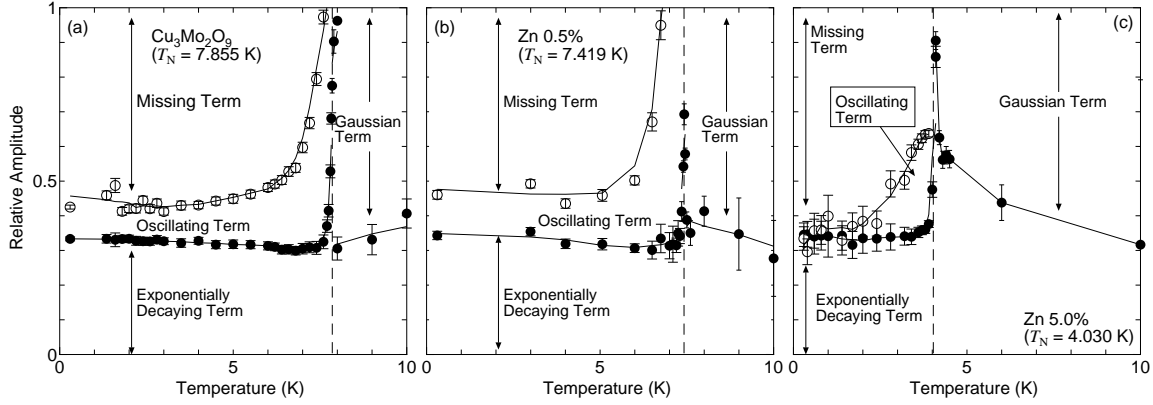}
\caption{
Amplitudes of $A_{\rm L}$ (closed circles) below and above $T_{\rm N}$ 
and $A_{\rm L}+A_{\rm O}$ (open ones) below $T_{\rm N}$ in 
(a) Cu$_{3}$Mo$_{2}$O$_{9}$, (b) the Zn-0.5\% doped sample, 
and (c) the Zn-5.0\% doped one.
The dashed lines denote $T_{\rm N}$s.
The curves denoting 
the relative amplitudes in these diagrams are 
only eye guides generated by the smoothed data.
}
\label{Amplitude}
\end{figure}

For $A(t)$ below $T_{\rm N}$ in the Zn-5.0\% sample, 
the widely distributing internal fields 
in the O term are reproduced by a Gaussian distribution, 
as shown in Fig. \ref{CorrectedAsymmetry_003K}(b).
And then, the fitting function becomes 
\begin{equation}
A_{\rm LT-G}(t)
= A_{\rm O} \cos{\left[\gamma H_{\rm int} t + \phi^{\mathstrut} \right]} 
            \exp{\left[ - \left( \sigma_{\rm O} t \right)^{2} \right]}
+ A_{\rm L} \exp{\left[ - \lambda_{\rm L} t^{\mathstrut} \right]} \ ,  
\end{equation}
with the factor $A_{\rm M} \equiv 1-A_{\rm O} - A_{\rm L}$.
The relative amplitude-temperature diagrams are given 
in Figs. \ref{Amplitude}(a)-\ref{Amplitude}(c).
Here the critical temperature is optimized 
so that the critical exponent of the internal field becomes 0.5.
As will be mentioned later, 
this value is predicted by using the molecular-field theory.
The $T_{\rm N}$s in all the samples 
obtained from the $\mu$SR measurements are consistent 
with the ones obtained from magnetization.

\begin{center}
\begin{figure}[h]
\includegraphics[scale=0.44]{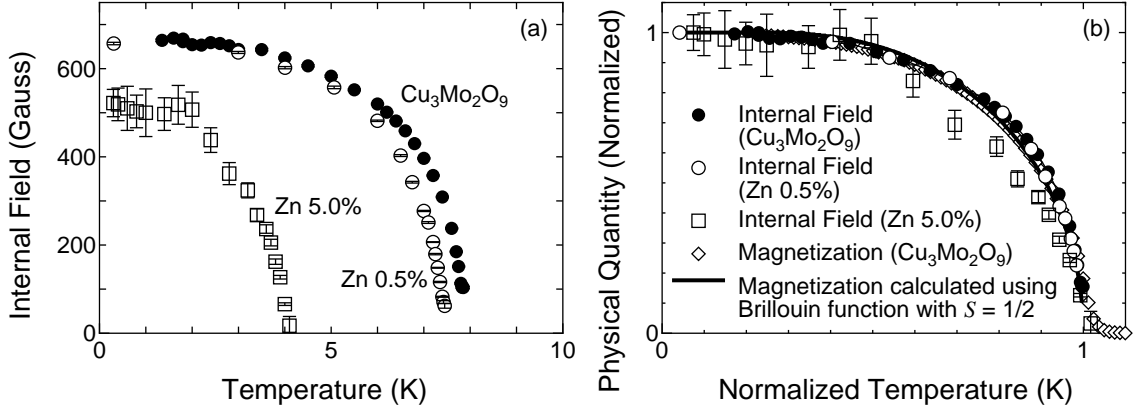}
\caption{(a) Internal field 
in Cu$_{3}$Mo$_{2}$O$_{9}$ (closed circles), 
the Zn-0.5\% (open circles) sample, 
and the Zn-5.0\% one (squares).
(b) The normalized internal fields 
with the same symbols as in (a) 
and the normalized magnetization (diamonds).
The magnetization calculated using the Brillouin function 
is shown by the bold curve.
All the physical quantities are normalized by the values 
at the lowest temperatures.}
\label{OrderParameters}
\centering
\end{figure}
\end{center}

Finally, in all the samples, 
we measured $A(t)$ at several temperatures 
and obtained $H_{\rm int}$ as functions of temperature.
The results are given in Fig. \ref{OrderParameters}(a).
Because these temperature dependences are similar to one another, 
we normalized the abscissa by $T_{\rm N}$ 
and the ordinate by the values at the lowest temperature.
The results are shown in Fig. \ref{OrderParameters}(b).

\section{Discussion}
\subsection{Magnetic Order Parameter}
At temperatures just below $T_{\rm N}$, 
$A_{\rm G}$ disappears completely and $A_{\rm O}$ increases.
These are due to the formation of 
slightly canted AFM order.
The $\mu$SR spectra 
in Cu$_{3}$Mo$_{2}$O$_{9}$ and the Zn-0.5\% doped sample 
have almost similar temperature dependence.
It is shown 
in the oscillating spectra [Fig. 3(a)]
and the distributing internal fields [Fig. 3(b)].
As shown in Fig. 5(b), 
the temperature dependences of 
internal fields in these samples are well scaled into one line.
On the other hand, the temperature dependences of the magnetization 
in these two systems are very different: 
In the Zn-0.5\% doped sample, 
the increase of $M_{a}$ caused by 
the spin canting under the magnetic field is not observed.
These indicate that 
the oscillation in $\mu$SR spectrum 
reflects the ferromagnetic long-range order in each sublattice 
which is the order parameter of AFM order.
Indeed, the temperature dependence of the internal fields in 
these two samples are well reproduced by 
the temperature dependence of the spontaneous magnetization 
in $S$ = 1/2 {\it ferromagnet} 
calculated using the Brillouin function \cite{Kittel}.

In case of the Zn-5.0\% doped sample, as shown in Fig. 3(b), 
the distribution of the internal field is very broad 
and its structure is more complicated.
As shown in Fig. 5(b), 
the normalized internal field in this sample 
is slightly different from the Brillouin function.
Moreover, the amplitude-temperature diagram in Fig. 4(c) 
does not seem to have similar temperature dependence 
when we compare it to those in Figs. 4(a) and 4(b).
These results strongly suggest the QPT induced by impurity doping; 
the system probably has a different magnetic structure.
The widely distributed internal field in Fig. 3(b) 
suggests a formation of magnetic superlattice below $T_{\rm N}$.
To discuss further, 
the magnetic structure determination 
in the Zn-5.0\% doped sample is necessary.

\subsection{Origin of Missing Term}
Because of finite pulse width in the pulsed $\mu$SR measurement, 
high-frequency signals are smeared out 
when the inversed frequency and the pulse width are comparable.
And then, there is a sensitivity limit in $H_{\rm int}$ (or in precession frequency).
$H_{\rm int}$ in the present work ($\sim$ 650 G) 
almost reaches this limit.
And then, 
the oscillating term at the low temperatures 
is probably underestimated.
To discuss more precisely, 
we need to consider the frequency dependence of 
the sensitivity seriously 
or to measure $\mu$SR spectra by using continuous muon source.
Both of them are beyond the scope of this paper.

\subsection{Muon Stopping Sites}
Figures 4(a)-4(c) show that 
the relative amplitude of the L term 
well below $T_{\rm N}$ is slightly larger than 0.3, 
which is almost independent of Zn concentration.
This term is a constant term coupled with the dynamical spin fluctuations.
Qualitatively, in all the samples, 
$A_{\rm L}$ is almost independent of temperature well below $T_{\rm N}$, 
has a peak around $T_{\rm N}$, 
and becomes about 0.3 again well above $T_{\rm N}$.
The value of $A_{\rm L} \sim 1/3$ reminds us 
the case of ferromagnetic polycrystal 
at zero magnetic field: 
The muon spin rotation spectrum oscillates 
around a constant asymmetry $A/3$ 
with an amplitude of $2A/3$, 
where $A$ is an initial asymmetry.
For another example of the value of 1/3, as is well known, 
the value of Kubo-Toyabe function reaches 1/3
without considering the dynamical decaying effects.
As an analogy to the value of 1/3 appearing in these famous cases, 
one might think that the value of 1/3 in the present case suggests something special.
However, the value of 1/3 in these cases 
is introduced through the powder average, 
not in the present case.
Moreover, 
the observed spectrum contains $A_{\rm L} \sim 1/3$ 
even in the paramagnetic phase:
A dynamically fluctuating internal field works 
on a third of all the muons in crystal.
The possible origins of this internal field 
are the phase separation effect 
and/or the two (or more than two) muon stopping sites.

\section{Conclusive Remarks}
We measured $\mu$SR spectra in Cu$_{3}$Mo$_{2}$O$_{9}$ 
and the lightly and the heavily Zn-doped samples.
We found that the order parameter of 
the phase transition at $T_{\rm N}$ was 
the sublattice magnetization which was $\mu$SR detectable.
This statement was confirmed 
in cases of Cu$_{3}$Mo$_{2}$O$_{9}$ and the lightly Zn-doped sample.
In the heavily Zn-doped sample, 
the $\mu$SR spectrum strongly suggests a formation of magnetic superlattice, 
which should be confirmed by using neutron diffraction.

Finally, as a proceedings paper of USM2013, 
we propose an application of ultra slow muon microscope:
the $\mu$SR measurement under transverse electric field 
to study the cross correlation effects in multiferroic material.
In conventional setting of $\mu$SR measurements, 
the lateral resolution (about 30 mm in diameter) 
and the muon penetration depth 
(depending on the kinetic energy of muon beam) 
cannot be controlled. 
And then, we could not control 
the relative direction 
between the muon's polarization direction 
and the electric field.
With higher lateral resolution and 
controllable muon energy due to reacceleration, 
the muon beam can be led into the plate-shape single crystal 
from the edge side, which expands the degree of freedom of 
experimental configurations.
And then, we can detect the magnetic order 
induced by an electric field more efficiently.
Of course, the bending of beam axis 
due to the static electric field working on the charge of muon 
may cause a problem, which will be solved by controlling 
the muon energy.

\section*{Acknowledgements}
This work was partly supported by 
a Grant-in-Aid for Scientific Research (C) (No. 22540350) 
of The Ministry of Education, Culture, Sports, Science, and Technology, Japan.

\end{document}